\documentclass{iau} 
\usepackage{graphicx}

\title[Multiwavelength morphological study] 
{Multiwavelength morphological study of \\ active galaxies}

\author[Betelehem Bilata-Woldeyes et al.]   
{Betelehem Bilata-Woldeyes$^{1,2}$,
Mirjana Povi\'c$^{2,3}$, Zeleke Beyoro-Amado$^2$, Tilahun Getachew-Woreta$^2$, Shimeles Terefe$^2$}

\affiliation{$^1$Debre Berhan University (DBU), Debre Berhan, Ethiopia; $^2$Ethiopian Space Science and Technology Institute (ESSTI), Addis Ababa, Ethiopia; $^3$Instituto de Astrof\'isica de Andaluc\'ia (IAA-CSIC), Granada, Spain}

\pubyear{2019}
\volume{356}  
\jname{Nuclear Activity in Galaxies Across Cosmic Time} 
\editors{M. Povi\'c, P. Marziani, J. Masegosa, H. Netzer,\\ S. H. Negu, \&
	S. B. Tessema, eds.}
\begin{document}

\maketitle

\begin{abstract}  
Studying the morphology of a large sample of active galaxies at different wavelengths and comparing it with active galactic nuclei (AGN)
properties, such as black hole mass (M$_{BH}$) and Eddington ratio ($\lambda_{Edd}$), can help us in understanding better the connection between AGN and their host galaxies and the role of nuclear activity in galaxy formation and evolution. By using the BAT-SWIFT hard X-ray public data and by extracting those parameters measured for AGN and by using other public catalogues for parameters such as stellar mass (M$_{*}$), star formation rate (SFR), bolometric luminosity (L$_{bol}$), etc., we studied the multiwavelength morphological properties of host galaxies of ultra-hard X-ray detected AGN and their correlation with other AGN properties. We found that ultra hard X-ray detected AGN can be hosted by all morphological types, but in larger fractions (42\%) they seem to be hosted by spirals in optical, to be quiet in radio, and to have compact morphologies in X-rays. When comparing morphologies with other galaxy properties, we found that
ultra hard X-ray detected AGN follow previously obtained relations. On the SFR vs. stellar mass diagram, we found that although the majority of sources are located below the main sequence (MS) of star formation (SF), still non-negligible number of sources, with diverse morphologies, is located on and/or above the MS, suggesting that AGN feedback might have more complex influence on the SF in galaxies than simply quenching it, as it was suggested in some of previous studies. 
\keywords{active galaxies, morphology, multiwavelength study, star formation rate}
\end{abstract}

\firstsection 
\section{Introduction}

Morphology is one of the key parameters for understanding the whole picture of galaxy formation and evolution along cosmic time 
(\cite[Conselice 2014]{conselice2014evolution}). It gives us an important information about galaxy structure and how other parameters 
such as environment, interactions and mergers, nuclear activity in galaxies, interstellar medium (ISM), etc., affect morphology and 
vice-versa. It also has a connection with other galaxy properties such as M$_*$ (\cite[Brinchmann \& Ellis 2000]{brinchmann2000mass}), 
SFR (\cite[Poggianti et al. 2008]{poggianti2008relation}), M$_{BH}$ (\cite[Schawinski et al. 2010]{schawinski2010galaxy}), etc. 

In addition, active galaxies, having active galactic nuclei (AGN) in their centers, emit radiation at different parts of 
electromagnetic spectrum (EMS). Different studies have been done regarding the connection between AGN and their host galaxies 
leading to different results (very often not very consistent). Taking into account X-ray data it has been suggested that majority 
of AGN reside in the green valley and could be a transitional population of galaxies (\cite[Povi\'c et al. 2009a]{povic2009otelo}, 
\cite[b]{povic2009anticorrelation}, \cite[2012]{povic2012agn}), moving from disk-dominated to bulge-dominated 
(\cite[Gabor et al. 2009]{gabor2009active}). Optical spectroscopic results suggested the same (e.g., 
\cite[Leslie et al. 2016]{leslie2016quenching}). However, other works in mid-infrared (MIR) and radio found AGN to be 
located in the blue cloud and red sequence, respectively (e.g., \cite[Hickox et al. 2009]{hickox2009host}). Most of 
morphological studies, at both lower and higher redshifts suggested that AGN tend to be hosted by massive elliptical 
or bulge-dominated galaxies (\cite[Kauffmann et al. 2003]{kauffmann2003host}, \cite[Povi\'c et al. 2012]{povic2012agn}), 
although the nearby Seyfert galaxies were found predominately in spirals (\cite[Ho 2008]{ho2008nuclear}). Being one of the 
key parameters, morphological properties of AGN are still not well known, and in particular there is a lack of multiwavelength 
morphological studies of active galaxies. 

We went for the first time through the multiwavelength morphological analysis in optical, radio, and X-rays of the ultra-hard 
X-ray AGN host galaxies in the BAT AGN spectroscopic survey (BASS). We tried to understand better the morphological properties 
of active galaxies at different wavelengths, and also their connection with supermassive black hole (SMBH) mass, Eddington ratio 
($\lambda_{Edd}$), bolometric luminosity (L$_{bol}$), SFR, and stellar mass (M$_*$). 

\section{Data}

Swift/BAT is an all-sky ultra-hard X-ray survey in 14 - 195 keV energy range. It is the only ultra hard X-ray survey of the whole sky  
(\cite[Ricci et al. 2017]{ricci2017bat}, \cite[Koss et al. 2017]{koss2017bat}). We used the BASS database 70-month Swift BAT 
all-sky catalogue with 1210 sources (\cite[Baumgartner et al. 2013]{baumgartner201370}). Out of these, we found optical 
photometric data for 640, 468, and 317 galaxies in optical, radio, and X-rays, using the Sloan Digital Sky Survey (SDSS) 
Data release 14 (DR14), FIRST and NVSS, and XMM-Newton, respectively. 

\section{Analysis and Results}

We classified galaxies visually in optical, radio and X-rays. We classified 45\% (290/640) of galaxies in  
optical (as elliptical, spiral, irregular, or peculiar), while other 55\% (350/640) of galaxies stayed unclassified due to the poor 
resolution data and/or edge-on sources. In radio we classified 84\% (394/468) of galaxies as radio-loud (RL) or radio-quiet (RQ), 
while other 16\% (74/468) are uncertain. Finally, we classified 99\% of X-ray sources as either compact or extended, while only 1\% 
(3/317) stayed uncertain. We found that ultra-hard X-ray AGN can be hosted by different morphologies, but mainly by spiral galaxies 
in optical, RQ in radio, and compact in X-rays.

We compared our morphological classification obtained at different wavelengths. We found 48\% (174/361) of sources with both optical 
and radio classifications, where the majority of sources (52\%) are spiral and RQ. With both optical and X-ray morphologies we found 
52\% (139/265) of sources, where again 55\% of sources are spiral in optical and compact in X-rays. Having both radio and X-ray 
classifications 85\% (172/202) of sources are found, with majority of sources being RQ (85\%) and compact in X-rays. We observed that 
there are very rare RL as well as extended X-ray galaxies in the BASS survey. 55\% (97/197) of sources have classifications in all 
three parts of EMS, of those 42\% are spiral in optical, RQ in radio, and compact in X-rays.

By studying the correlation between multiwavelength morphology and other AGN properties we found early-type (ET) galaxies to have 
more massive M$_{BH}$, higher L$_{bol}$, and slightly higher M$_*$ than late-type (LT) galaxies, while galaxies classified as 
LT have slightly higher accretion rate as shown in Figure 1 (as raised in some of previous studies, e.g., 
\cite[Povi\'c et al. 2009a]{povic2009otelo}, \cite[Fanidakis et al. 2010]{fanidakis2010grand}, 
\cite[Koss et al. 2011]{koss2011host}). For the radio and X-ray sources, galaxies classified as 
RL have a bit larger M$_{BH}$, higher luminosity, and higher M$_*$ than RQ sources and extended X-ray sources have 
higher $\lambda_{Edd}$, higher M$_*$ and lower luminosity than compact X-ray sources.\\

\begin{figure}[h!]
\centering
  \includegraphics[width=2.3in, height=3.09cm]{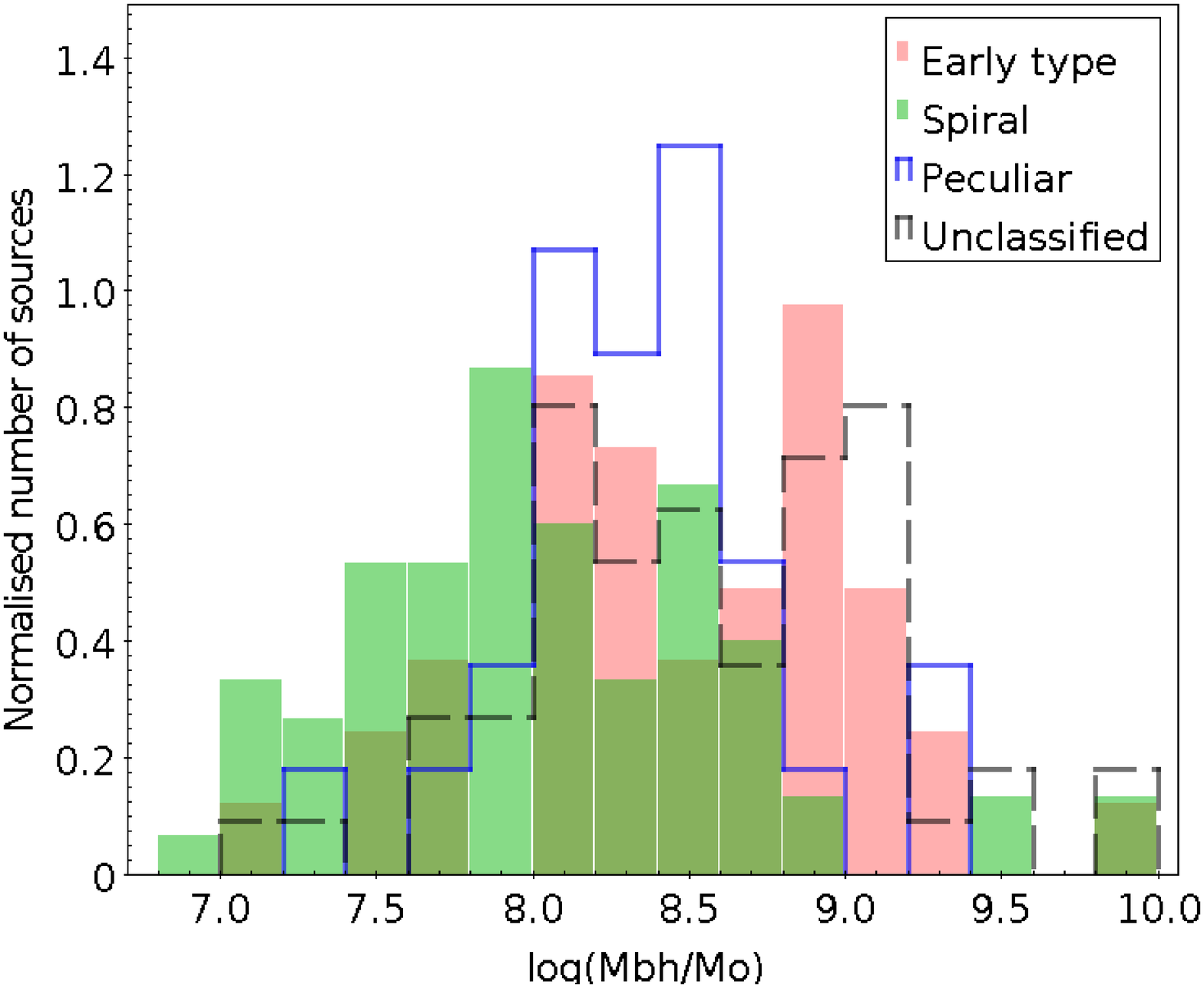} 
  \includegraphics[width=2.3in, height=3.09cm]{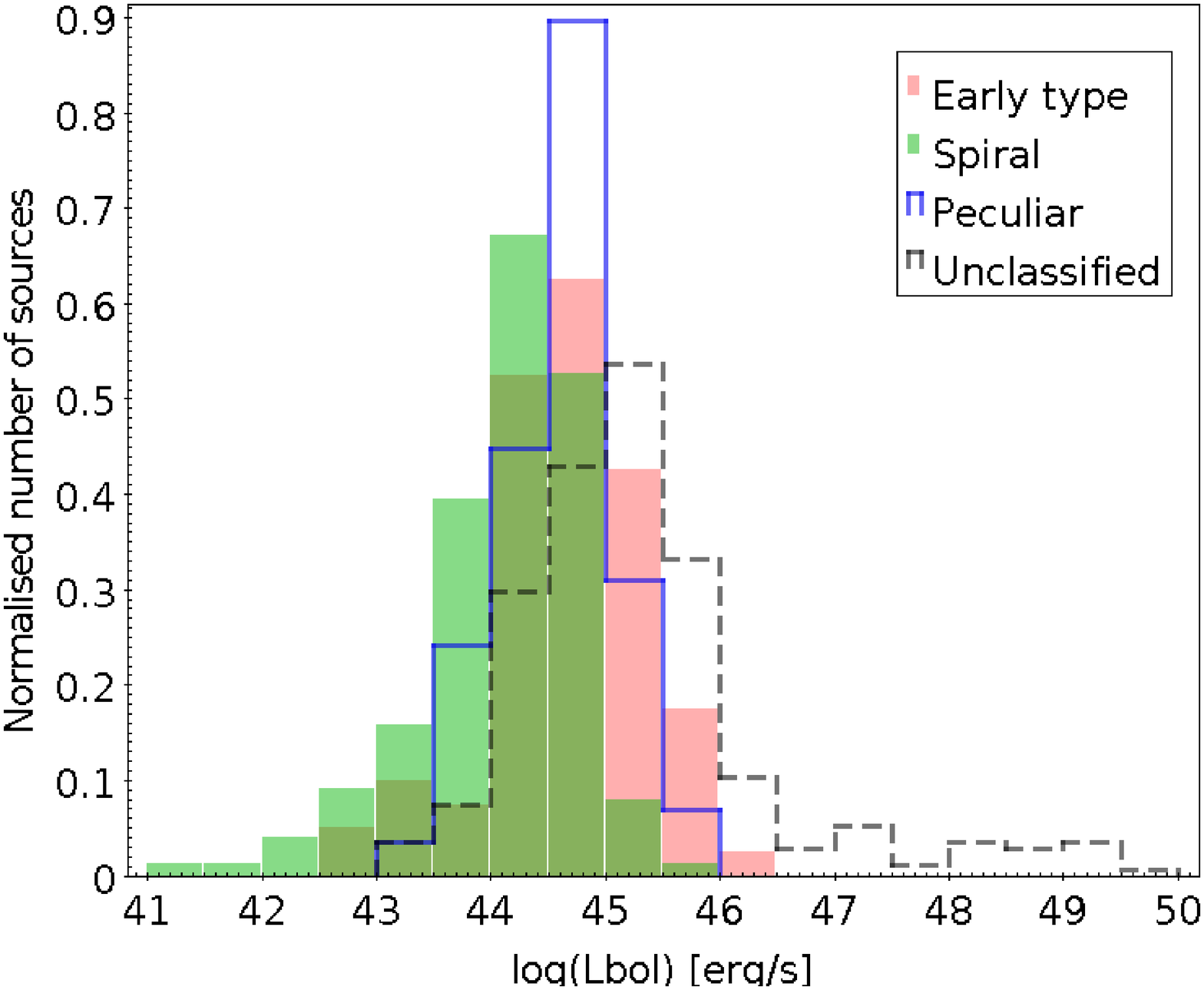} 
  \includegraphics[width=2.3in, height=3.09cm]{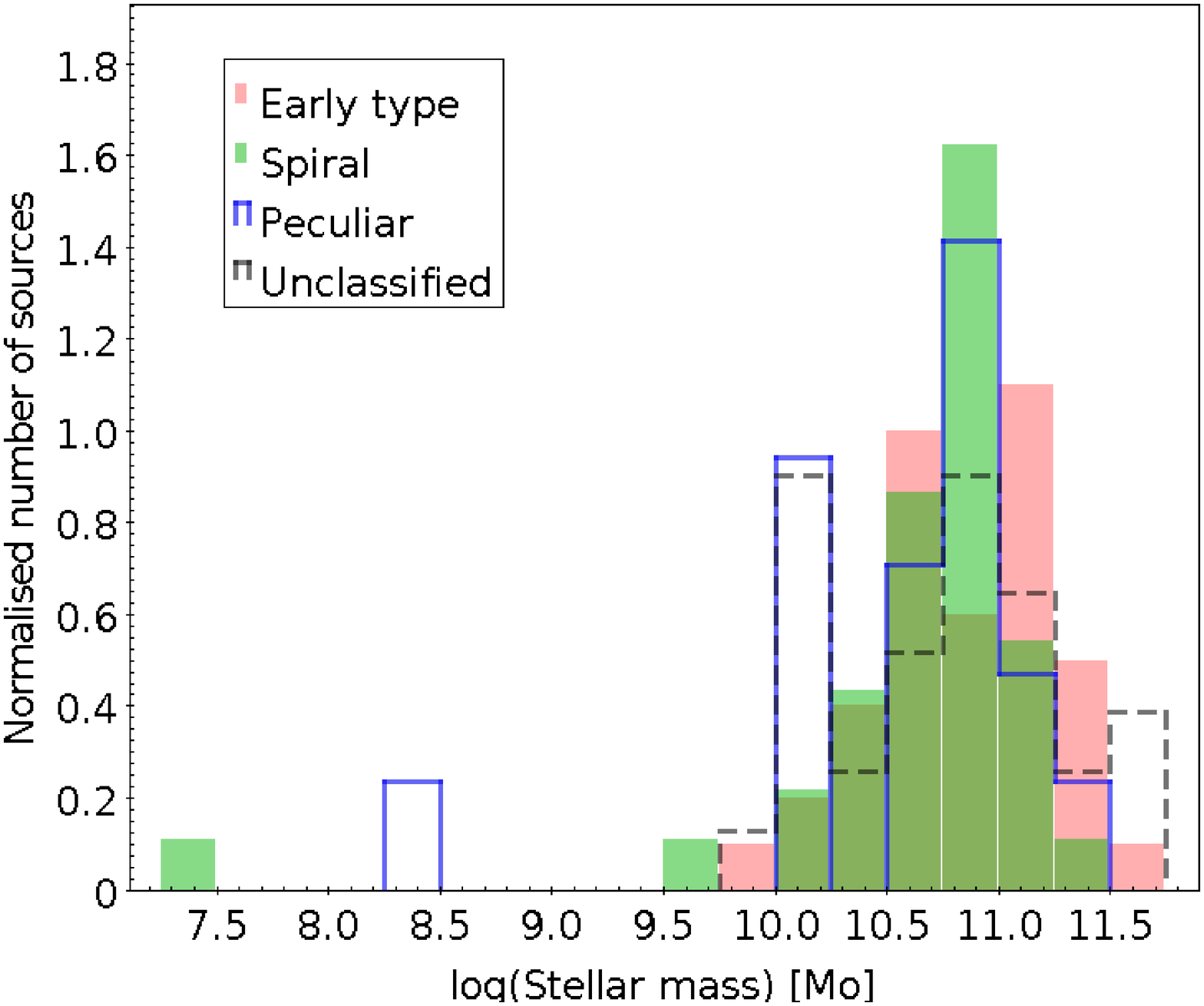} 
  \includegraphics[width=2.3in, height=3.09cm]{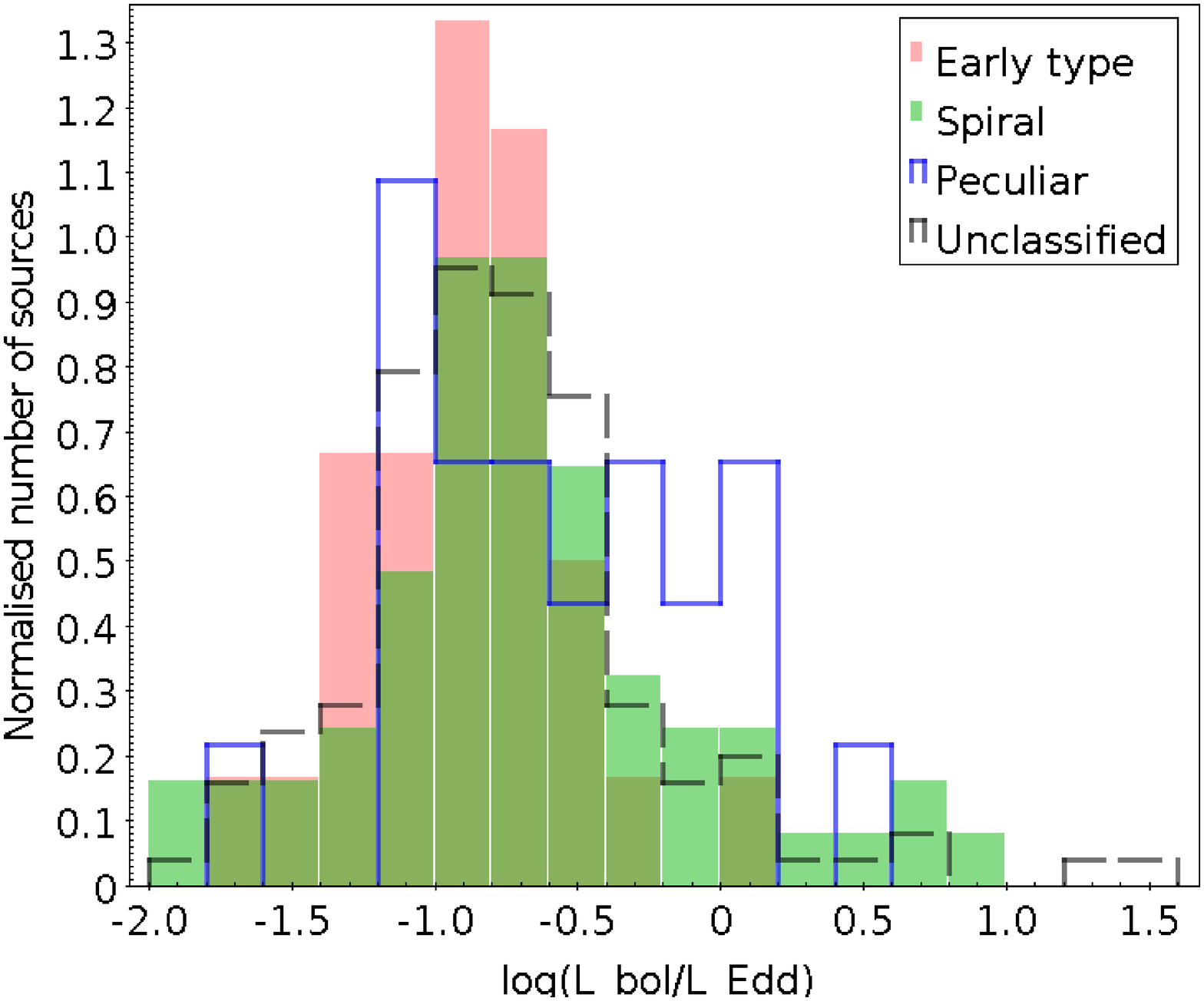} 
 \caption{From top to bottom and from left to right: Distribution of M$_{BH}$, L$_{bol}$, M$_*$, and $\lambda_{Edd}$ of optically classified sources. Different color of the histogram is related with different morphological type, as indicated on each plot.} 
 \label{fig.p}
\end{figure}

We studied the location of our sources on the M$_*$ - SFR diagram, where for the main sequence (MS) of star formation (SF) we used results from \cite[Elbaz et al. 2007]{elbaz2007reversal}. In general we found that ultra-hard X-ray detected AGN can be on, below, and above the MS of SF. Although we found higher population to be located below the MS, having galaxies on and above the MS might suggest that not necessarily AGN are responsible for SF quenching. 
\vspace{-0.5 cm}
\section*{Acknowledgements}

We thank the financial support from the DBU, Ethiopian Ministry of Higher Education and the EORC under ESSTI.  MP also acknowledges the support from the Spanish Ministry of Science, Innovation and Universities (MICIU) through projects AYA2013-42227-P and AYA2016-76682C3-1-P. Finally, we thank to the BASS collaboration for 
making there data available to public.


\end{document}